\documentclass[a4paper]{jpconf}

\usepackage[subtle]{savetrees}

\usepackage{float} 

\usepackage{graphicx}

\usepackage{amsmath, amsthm, amssymb, amsfonts}

\usepackage{lineno}

\usepackage{subfigure}
\usepackage{bm}
\usepackage[percent]{overpic}
\usepackage{color}

\usepackage{hyperref}
\usepackage[all]{hypcap}
\hypersetup{
    colorlinks,
    citecolor=blue,
    filecolor=blue,
    linkcolor=black,
    urlcolor=blue
}

\def\nn{\textit{n--n}}

\def\ptpt{{$p^{}_t$--$p^{}_t$}\ }

\def\nF{{n_{\rm F}^{}}}

\def\nB{{n_{\rm B}^{}}}

\def\eg{\eta_{\rm gap}}
\def\deta{\delta\eta}

\def\pt{p_{\rm T}}

\def 	\bcor{b_{\rm corr}}
\def 	\C2{C_{\rm 2}}

\def\av#1{\langle #1 \rangle}

\def\nn{n-n} 
\def\ptpt{\overline{p_{\rm T}}-\overline{p_{\rm T}}}

\def\sNN{\mbox{$\sqrt{s_{_{\rm NN}}}$}}

\def\pf{\overline{p_{\rm F}}}
\def\pb{\overline{p_{\rm B}}}

\usepackage[pscoord]{eso-pic}

\newcommand{\placetextbox}[3]{
  \setbox0=\hbox{#3}
  \AddToShipoutPictureFG*{
    \put(\LenToUnit{#1\paperwidth},\LenToUnit{#2\paperheight}){\vtop{{\null}\makebox[0pt][c]{#3}}}%
  }%
}%

\begin{document}

\vspace*{-4.5cm}

\title{Forward-backward correlations between mean transverse momenta in Pb--Pb collisions with ALICE}

\author{Igor Altsybeev$^1$  (for the ALICE collaboration)}
\address{$^1$ 
Saint-Petersburg State University, 7/9 Universitetskaya nab., St. Petersburg, 199034 Russia
}

\ead{i.altsybeev@spbu.ru}

\begin{abstract}
Forward-backward (FB) correlations are considered to be a powerful tool for the exploration 
of the early dynamics of hadronic interactions.
The FB correlation functions can be constructed from different observables calculated event-by-event  in two separated pseudorapidity regions.
We report measurements of event-by-event average transverse momentum correlations
for charged particles 
in two separated pseudorapidity regions in Pb--Pb collisions at $\sqrt{s_{\rm NN}}=2.76$ and 5.02 TeV recorded with ALICE at the LHC.
The event-by-event mean transverse momenta correlations are robust against volume fluctuations  
and thus the centrality determination methods,
which provides higher sensitivity to the properties of the initial state 
and  evolution of the medium created in  A--A collisions.
The strength of the FB correlation is calculated for
different centralities of the Pb--Pb collisions.
Results are compared with Monte Carlo event generators, such as HIJING and AMPT.
\end{abstract}

\section{Introduction}

\placetextbox{0.88}{0.075}{1}

Studies of Forward-Backward  (FB) 
correlations are performed
between observables in two  separated
pseudo-rapidity intervals $\Delta \eta_F$ and $\Delta \eta_B$, which are conventionally
referred to as forward 
 and backward 
windows.
The FB correlations  are created predominantly  at the early stages of the collision~\cite{dumitru} 
and arise in such initial state models as  Color Glass Condensate \cite{cgc} 
and String Fusion~\cite{string_fusion}.
The FB correlations are sensitive to event-by-event fluctuations
of number and properties of particle-emitting sources elongated in rapidity,
and in later stages of the system evolution the  correlations
can be modified by medium and final state effects. 

The strength of the FB correlation  is usually characterized by the correlation coefficient
$\bcor$, which is obtained from a linear regression analysis of the event-averaged quantity
measured  in the backward rapidity hemisphere ($\av{B}_F$) 
as a function of the quantity measured 
in the forward hemisphere ($F$): 

\begin{equation}
\label{linear_FB}
\av{B}_{F} = a + \bcor\cdot F \  .
\end{equation}

\noindent Alternatively, the  $\bcor$ can be determined 
using the Pearson correlation coefficient:

\begin{equation}
\label{bcor}
\bcor= \frac{\av{FB}-\av{F}\av{B}} {    \av{F^2} - \av{F}^2}\  ,
\end{equation}

\noindent where angular brackets denote averaging over events.
Different dynamical variables can be chosen in $F$ and $B$ windows in order to study correlations between them. 
In conventional FB measurements, multiplicities of charged particles  $\nF$ and $\nB$
within the windows are chosen.
We refer to this kind of FB correlations as $\nn$ correlations, and 
formulae \eqref{linear_FB} and \eqref{bcor}  for them as 

\begin{equation}
\label{linear}
\av{\nB}_\nF = a + \bcor^{\nn}\cdot\nF \  , 
\  \  \   \bcor^{\nn} = \frac{\av{\nB\nF}-\av{\nB}\av{\nF}} {\av{n_{\rm F}^2}-\av{\nF}^2_{}}  \  .
\end{equation}

\noindent The forward-backward  multiplicity correlations have been previously studied
in a large number of colliding systems, for instance, in ${\rm p}\overline{\rm p}$ \cite{ua5},
  pp  \cite{FB_pp}
and Au--Au \cite{STAR_FB} collisions.

Charged particle multiplicity  is an {\it extensive} quantity,
therefore the strength of the FB $\nn$ correlations is affected by 
the so-called ``volume fluctuations" (i.e., in Glauber-like models, by
event-by-event fluctuations of the number of 
participating nucleons),
which  complicates the interpretation of the experimental values  of this observable.
To  suppress  this contribution,
one may consider other, {\it intensive} observables within the 
observation windows.
In particular,  
the mean transverse momentum of particles in a given event 
can be determined 
within each of the $F$ and $B$  windows,
given by the expressions
$F\equiv 
\pf=\sum_{j=1}^{n_F}\pt^{(j)}/\nF$ 
and
$B\equiv 
\pb= \sum_{i=1}^{n_B}\pt^{(i)}/\nB$ \cite{ALICE_PPR_v2}.
Then the formulae  \eqref{linear_FB} and \eqref{bcor} for
the correlation strength are 
expressed  as 

\begin{equation}
\label{bcor_ptpt_denomF}
\av{\pf}_{\pb} = a +\bcor^{\ptpt}\cdot\pf \  , \  \  \   
\bcor^{\ptpt}
= \frac{\av{\pf\  \pb}-\av{\pf}\av{\pb}} {    \av{\pf^2} - \av{\pf}^2} \ .
\end{equation}

In this work, 
FB correlations between charged primary particles
have been measured with the ALICE detector in Pb--Pb collisions 
at $\sNN=$2.76 and 5.02 TeV.
We first show an analysis
of FB multiplicity correlations and after that present
results on FB correlations between mean-$\pt$.

\section{Forward-backward correlations between multiplicities }

\placetextbox{0.88}{0.075}{2}

In the ALICE setup,
charged 
particles are reconstructed using combined information from the Inner Tracking System (ITS) and the Time Projection Chamber (TPC). Both detectors are located inside the ALICE solenoid with a field of 0.5 T and have full azimuthal coverage for track reconstruction within a pseudo-rapidity window of $|\eta| < 0.8$  \cite{alice_performance}.
Centrality of Pb--Pb collisions is determined using the signals from the V0 detectors -- 
two forward scintillator arrays with coverage $-3.7<\eta<-1.7$ and $2.8<\eta<5.1$.
Alternatively,  centrality can be estimated using 
the signal from spectators in the Zero-Degree Calorimeters  coupled with 
the response from  a small electromagnetic calorimeter ZEM (ZDCvsZEM estimator), 
as well as 
using the number of clusters counted in the second layer of the Silicon Pixel Detector covering $|\eta| < 1.4$ (CL1 estimator) \cite{ALICE_centrality_PbPb}.
Centrality classes are defined as percentiles of the multiplicity 
distributions.

\begin{figure}[!b]
\centering
\subfigure[a][]
{
\begin{overpic}[width=0.48\textwidth]{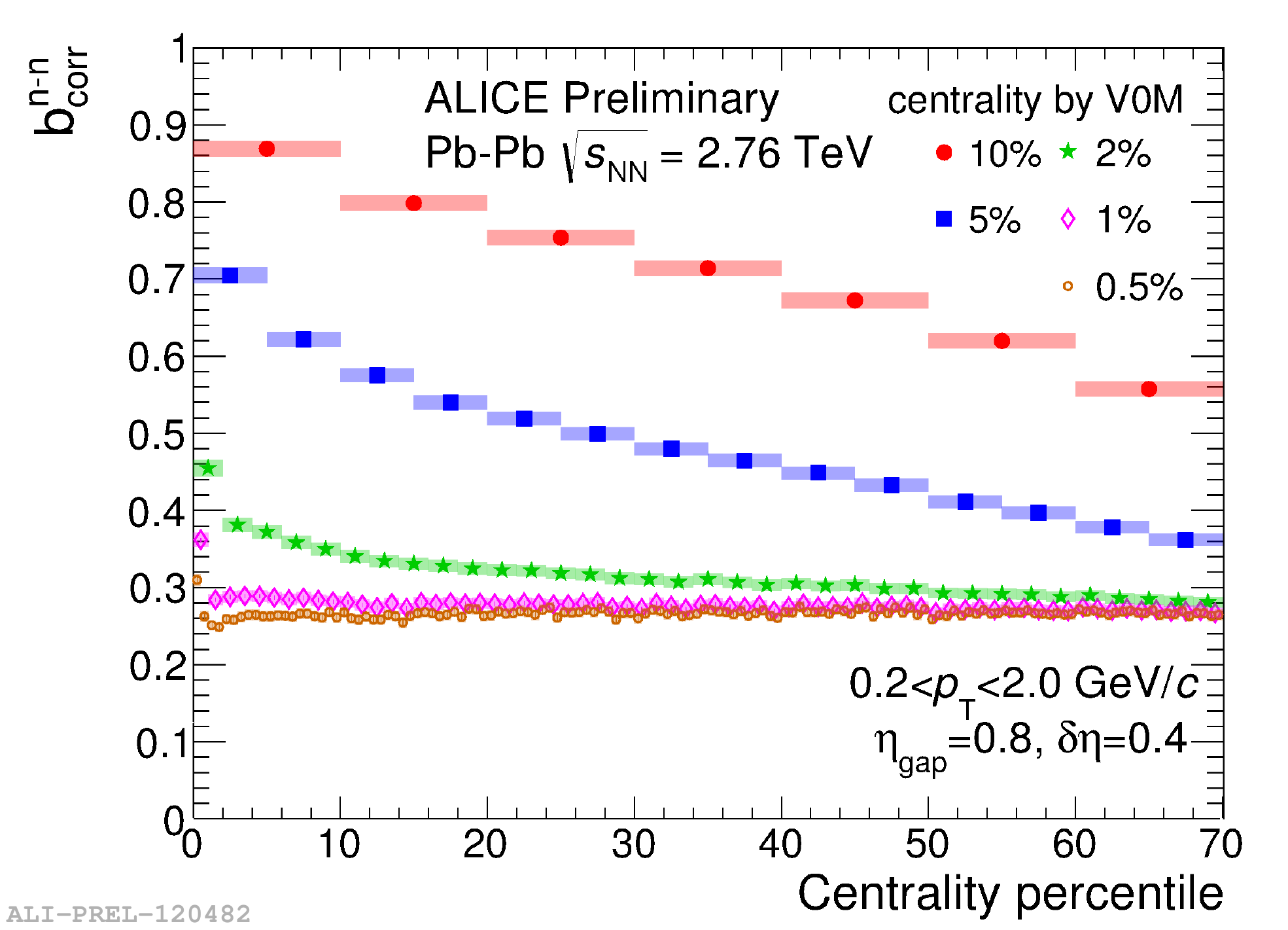}
\end{overpic}
}
\hspace{-0.4cm}
\subfigure[a][]
{
\begin{overpic}[width=0.48\textwidth]{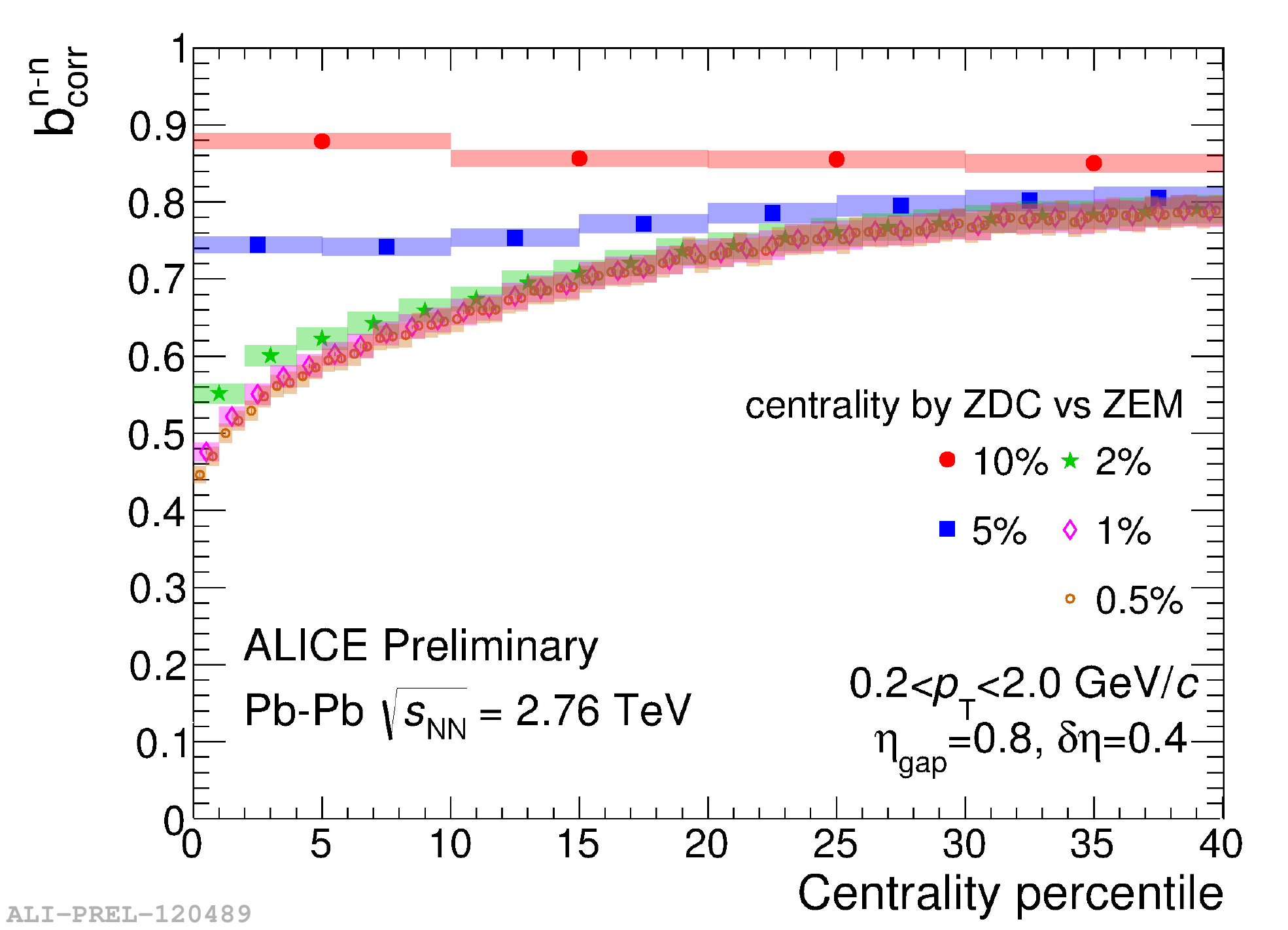}
\end{overpic}
}
\caption{
Strength of the FB multiplicity correlations 
as a function of centrality in Pb--Pb collisions at $\sNN=2.76$ TeV.
Centrality classes of different width are determined by the V0 (a) and by the ZDCvsZEM (b). Systematic uncertainties are shown as rectangles (widths correspond to the sizes of centrality classes),
statistical uncertainties are smaller than marker sizes.
}
\label{fig:for_approval_nn} 
\end{figure}

A pair of $\eta$ intervals chosen for this FB correlation analysis are (-0.8, -0.4) and (0.4, 0.8), 
which have a width $\deta=0.4$ and pseudorapidity separation 
 $\eg=0.8$. 
Such a separation allows short-range effects from (mini-)jets and resonance decays to be reduced.
A ``soft'' $\pt$-range of 0.2-2.0~GeV/$c$ was chosen for  the study. 
The numbers  of  Pb--Pb events selected for analysis 
are $12 \times 10^6$ at $\sNN=2.76$ TeV
and $49 \times 10^6$ at $5.02$ TeV.

Figure \ref{fig:for_approval_nn} shows the centrality dependence of the FB correlation strength $\bcor^{\nn}$ 
between multiplicities in the two windows.
The centrality estimator used for panel (a) is the V0 detector.
For wide centrality classes of 10\% width (red filled circles),  the correlation strength grows towards more central collisions, as it was observed by the STAR collaboration for Au--Au collisions~\cite{STAR_FB}. 
However,  for smaller 
class widths (5, 2, 1 and 0.5\%) the values of $\bcor^{\nn}$
drop and the centrality trend flattens, 
because the contribution from 
the volume fluctuations is suppressed for  narrower centrality classes.

Panel (b) shows values of $\bcor^{\nn}$
in classes of centrality determined by the ZDCvsZEM estimator.
It can be seen that the trends 
are very different in comparison with the V0-based results.
This is
because 
acceptance and resolution of the ZDCvsZEM estimator is distinct from those of the V0,
therefore, volume fluctuations inside centrality classes are different.
Moreover, a cross-correlation between ZDC with the central-barrel region is 
also not the same as in case of the V0, 
and  it is known also that resolution of 
the ZDC
worsens towards more peripheral collisions~\cite{ALICE_centrality_PbPb}. 
All these effects 
contribute to 
$\bcor^{\nn}$. 
Therefore, in view of a dramatic dependence of FB multiplicity correlation strength on centrality class determination, theoretical interpretation of the experimental results 
should be done with care.

\section{Forward-backward correlations between event-mean transverse momenta}

\placetextbox{0.88}{0.075}{3}

\begin{figure}[H] 
\centering
\subfigure[a][]
{
\begin{overpic}[width=0.36\textwidth]{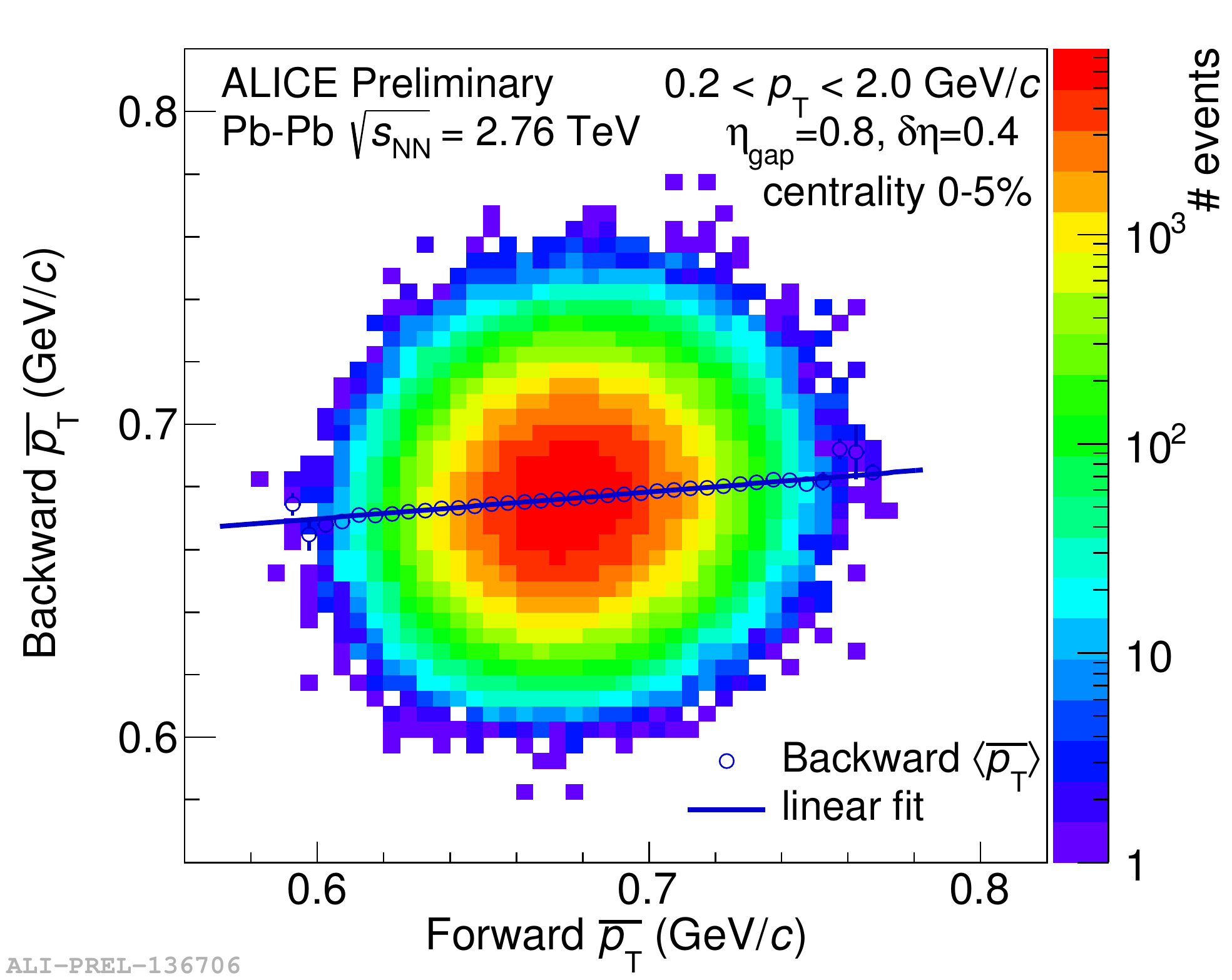}
\end{overpic}
}
\hspace{-0.5cm}
\subfigure[a][]
{
\begin{overpic}[width=0.31\textwidth]{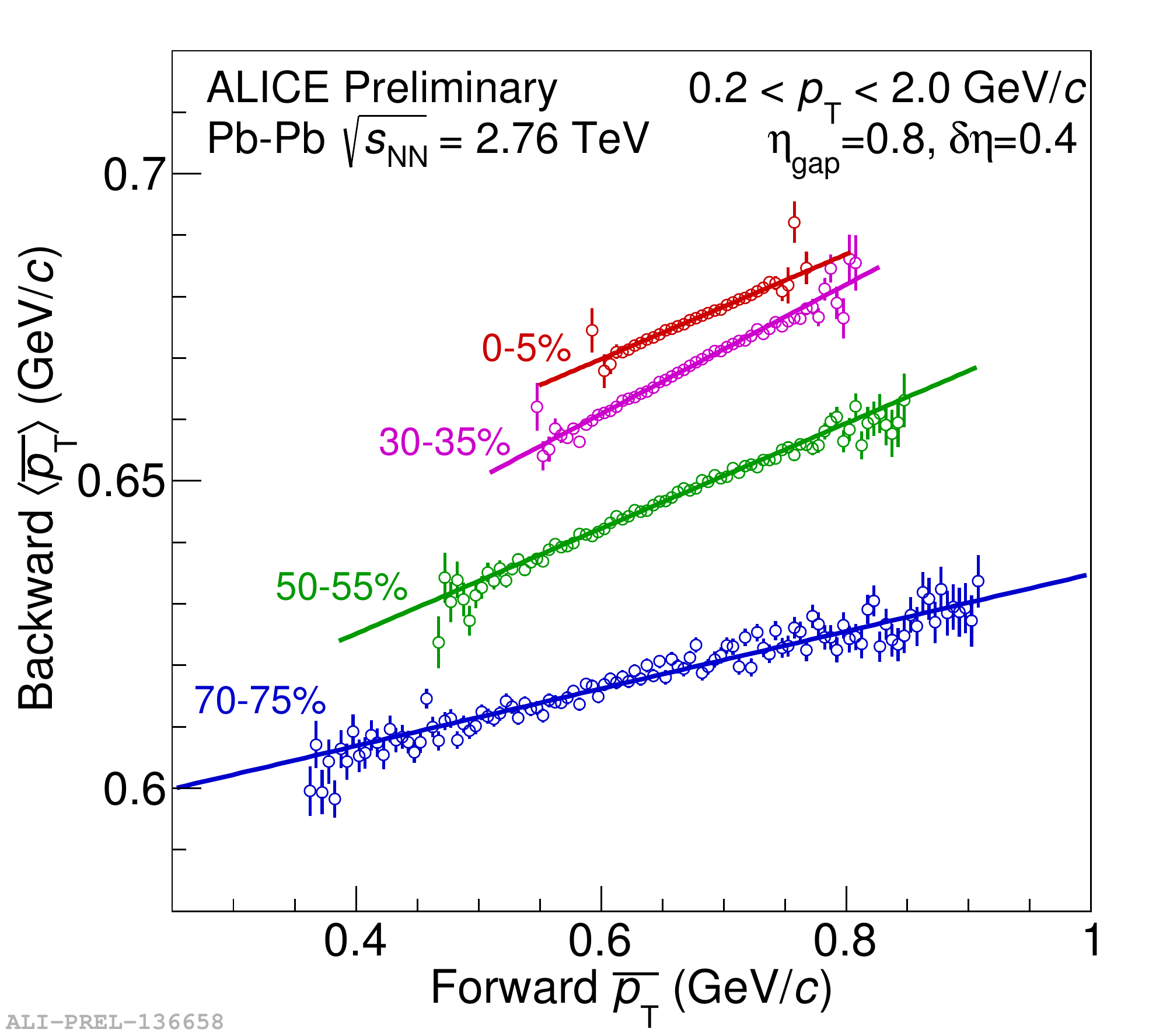}
\end{overpic}
}
\hspace{-0.6cm}
\subfigure[a][]
{
\begin{overpic}[width=0.31\textwidth]{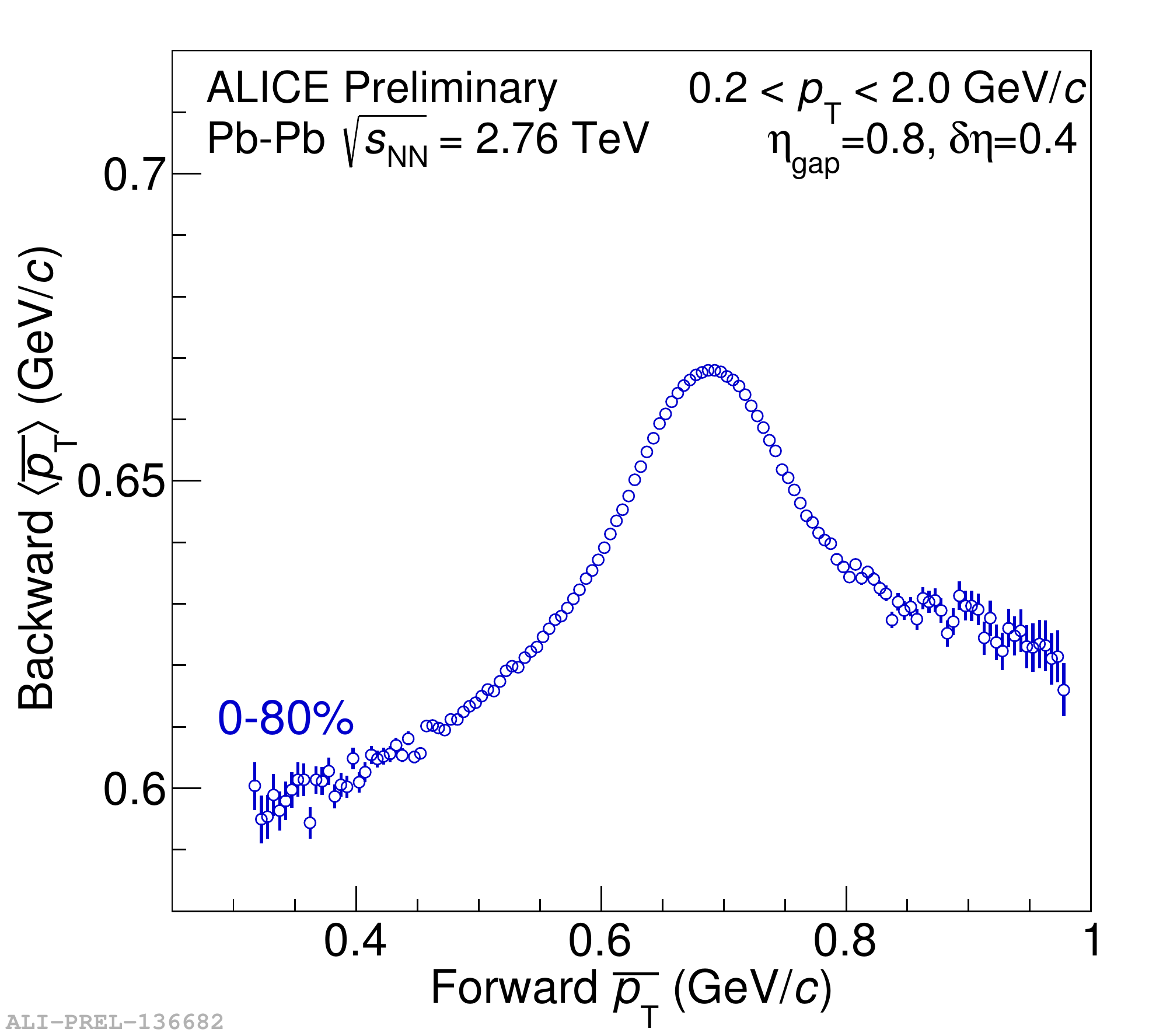}
\end{overpic}
}
\caption{
Mean $\pt$ in Backward window $vs$ mean $\pt$ in Forward window (with corresponding profile) in class 0--5\% (a),
profiles with linear fits for several centrality classes of 5\% width (b)
and profile for wide centrality class 0--80\% (c). Centrality is determined by the V0 detector.
}
\label{fig:HIST2D_FB_CORR} 
\end{figure}

Instead of multiplicities, correlations between event-mean $\pt$
have been studied
for the same FB window pair.
Figure \ref{fig:HIST2D_FB_CORR} (a)
shows an event-by-event distribution of $\pf$  versus $\pb$
for centrality class 0--5\% (centrality is determined by the V0 estimator).
For a linear regression analysis,
event-averaged values in the backward window
($\av{\pb}$) are calculated for each $\pf$ bin: 
panel (b) demonstrates this for several centrality classes of 5\% width.
One may note that  correlation functions are linear  in narrow centrality  classes, 
therefore each function can be quantified by the correlation
strength $\bcor^{\ptpt}$, which corresponds to the slope of the linear fit line. 
Note, that for too wide classes the linearity of the 
correlation functions may be broken: an extreme case is shown in 
panel (c) for the 0--80\% class.
Therefore, to interpret  the meaning of  $\bcor^{\ptpt}$,
it is important to look at the correlation functions themselves.

\vspace*{-0.3cm}

\vspace*{-0.3cm}

\begin{figure}[!t]
\centering
\subfigure[a][]
{
\begin{overpic}[width=0.48\textwidth]{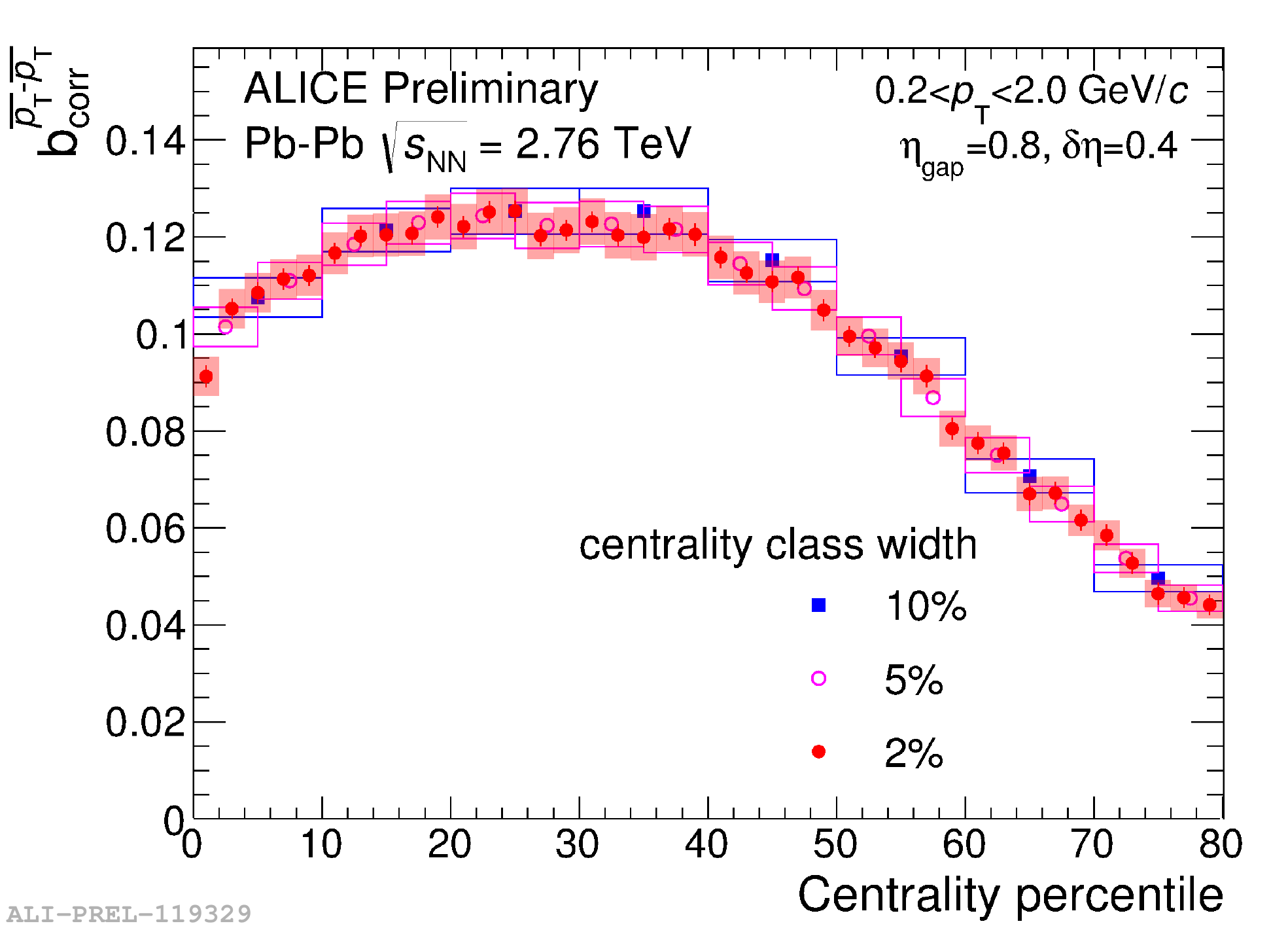} 
\end{overpic}
}
\hspace{-0.4cm}
\subfigure[a][]
{
\begin{overpic}[width=0.48\textwidth]{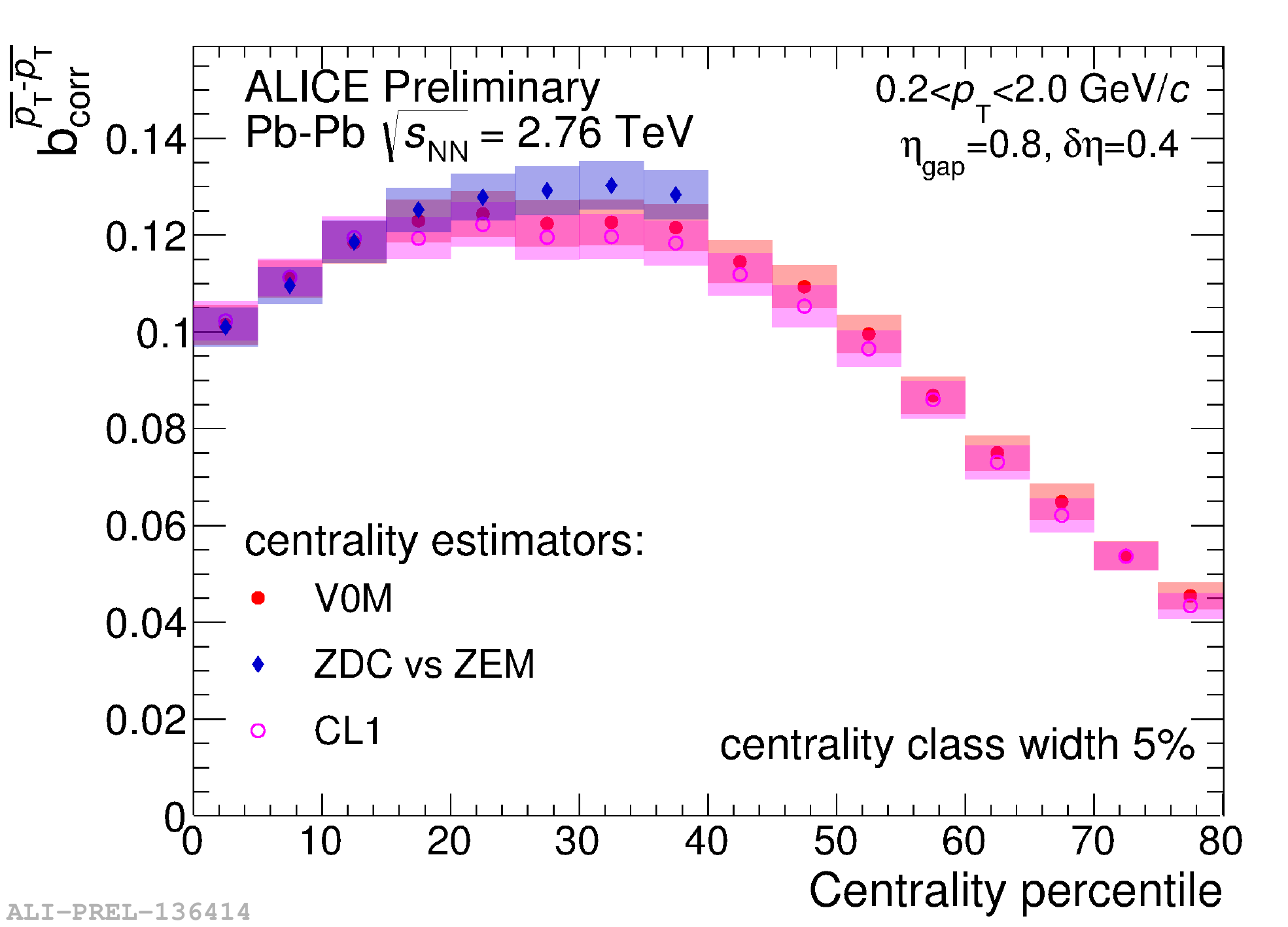}
\end{overpic}
}
\caption{
Dependence of $\bcor^{\ptpt}$  on centrality
for  classes of 10, 5 and 2\% widths,  determined by the V0 (a).
Results for centrality classes of 5\% width determined 
by the V0, ZDCvsZEM and CL1 estimators (b).
}
\label{fig:ptpt_centr_dep_cWidths} 
\end{figure}

\begin{figure}[!b]
\centering
\subfigure[a][]
{
\begin{overpic}[width=0.48\textwidth]{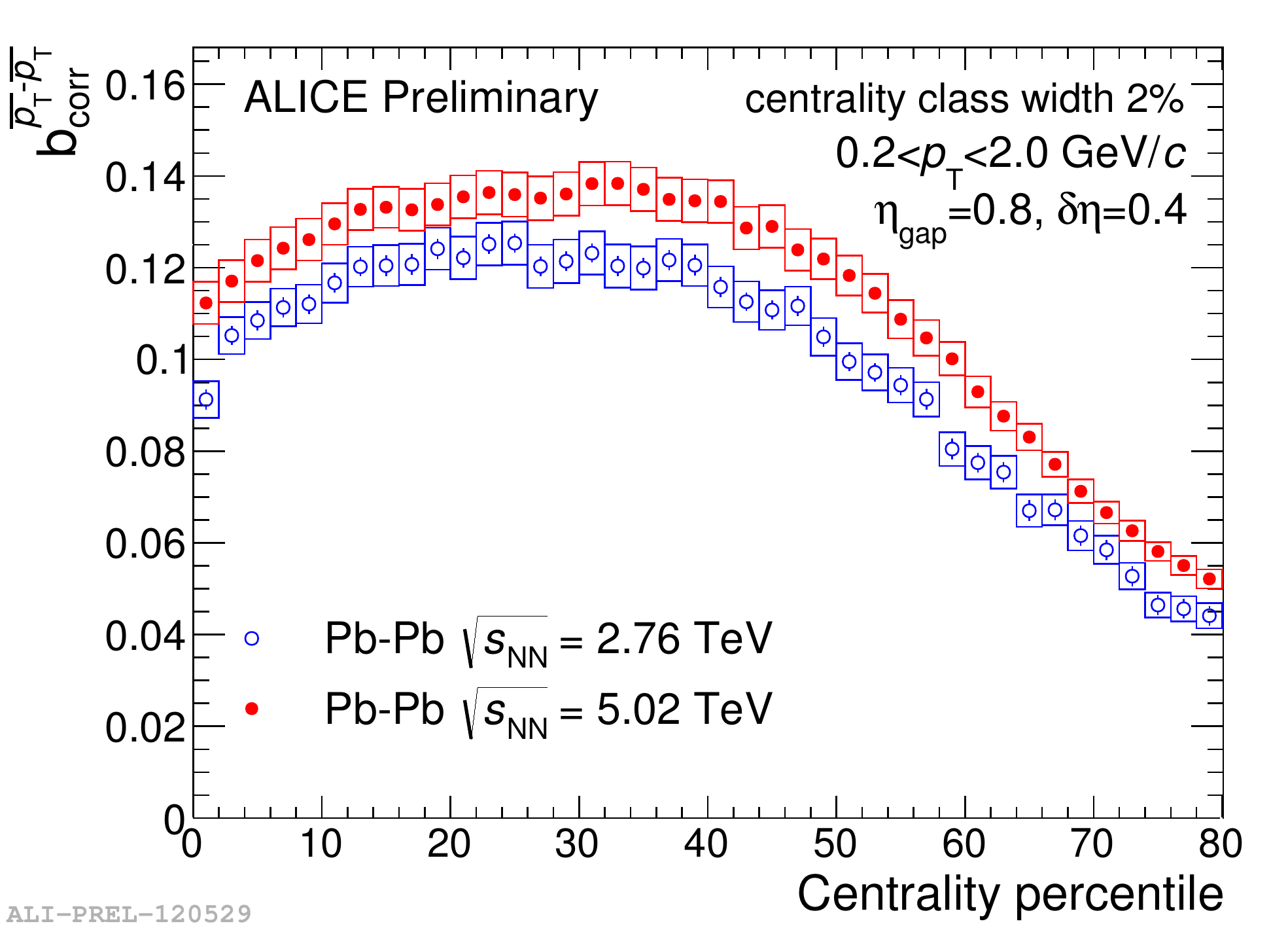}
\end{overpic}
}
\hspace{-0.4cm}
\subfigure[a][]
{
\begin{overpic}[width=0.48\textwidth]{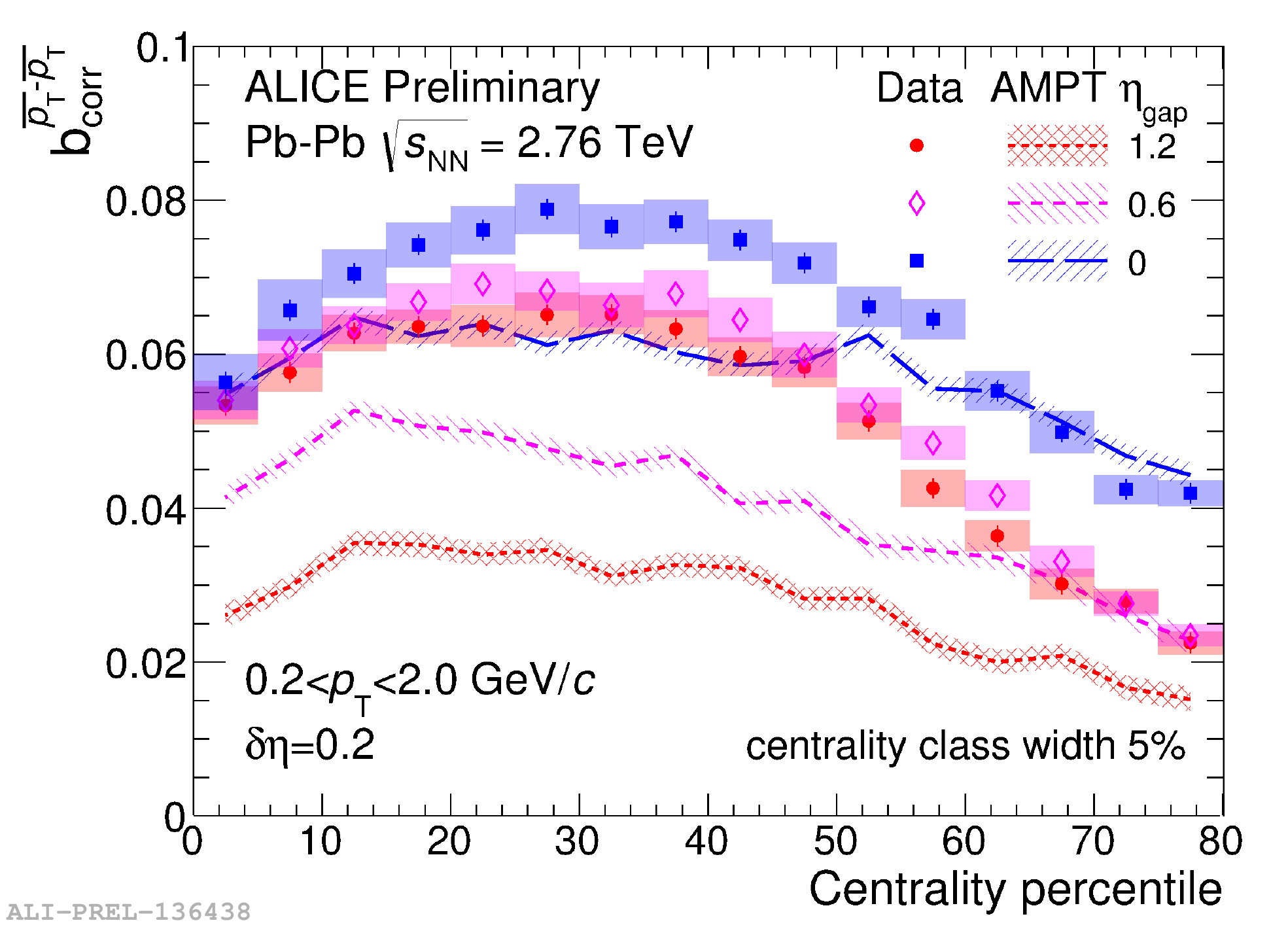}
\end{overpic}
}
\caption{ 
Centrality dependence of $\bcor^{\ptpt}$  (a) at two energies of Pb--Pb collisions $\sNN=$2.76 and 5.02 TeV,
and (b) for three $\eta$ gaps between $F$ and $B$ windows $\eg=$1.2, 0.6 and 0. 
Lines correspond to calculations with AMPT event generator.
Centrality by the V0 detector.
}
\label{fig:ptpt_eta_gaps_energy} 
\end{figure}

The centrality dependence of $\bcor^{\ptpt}$ is presented in 
Figure \ref{fig:ptpt_centr_dep_cWidths} (a).
Since mean-$\pt$ is an intensive observable, 
the results are independent of volume fluctuations 
and therefore are robust to changes of the centrality class width, if the classes are not too wide (points 
for 10, 5 and 2\% classes are shown).
Moreover, different centrality estimators also provide consistent results
(panel b). 
The small deviations in the centrality range 20--40\% for the ZDC-based results 
can be attributed to a reduced centrality resolution of the ZDC in this centrality range,
mentioned above.
The correlation strength $\bcor^{\ptpt}$ rises 
from peripheral to mid-central and
drops towards central collisions.
This characteristic shape persists also at $\sNN=5.02$ TeV energy of Pb--Pb collisions (Figure \ref{fig:ptpt_eta_gaps_energy}, a). 
Also, the same behavior is seen for windows of smaller width $\deta=0.2$
with different gaps between them (panel b): the results for $\eg=1.2$ and $0.6$ are on top 
of each other, while values for adjacent FB windows ($\eg=0$, blue squares) are slightly higher due to short-range correlations. The short-range contribution is most pronounced for peripheral events and decreases
towards central events.

The positive values of  $\bcor^{\ptpt}$ seen above
are related to the event-by-event fluctuations of the
event-averaged transverse momentum of  particles. 
The origin of these mean-$\pt$ fluctuations can be attributed, for example,
to event-by-event fluctuations of the initial size of the fireball,  
which is reflected in  pressure gradients 
at a later stage of a collision 
\cite{Bozek_Broniowsky_pt_flow_corrs}.
What is more difficult, however, 
is to capture correctly the {\it shape} of the centrality dependence of the $\bcor^{\ptpt}$.
Qualitatively, 
the same shape was obtained in the Monte Carlo 
implementation of the 
string fusion model \cite{MC_SFM},
where fluctuations of   initial densities  provide different 
patterns of overlapping strings, 
and changes of  tension of fused strings  
affect $\pt$ of particles emitted when the strings break.

\placetextbox{0.88}{0.075}{4}

\begin{figure}[H]
\centering
\begin{overpic}[width=0.6\textwidth]{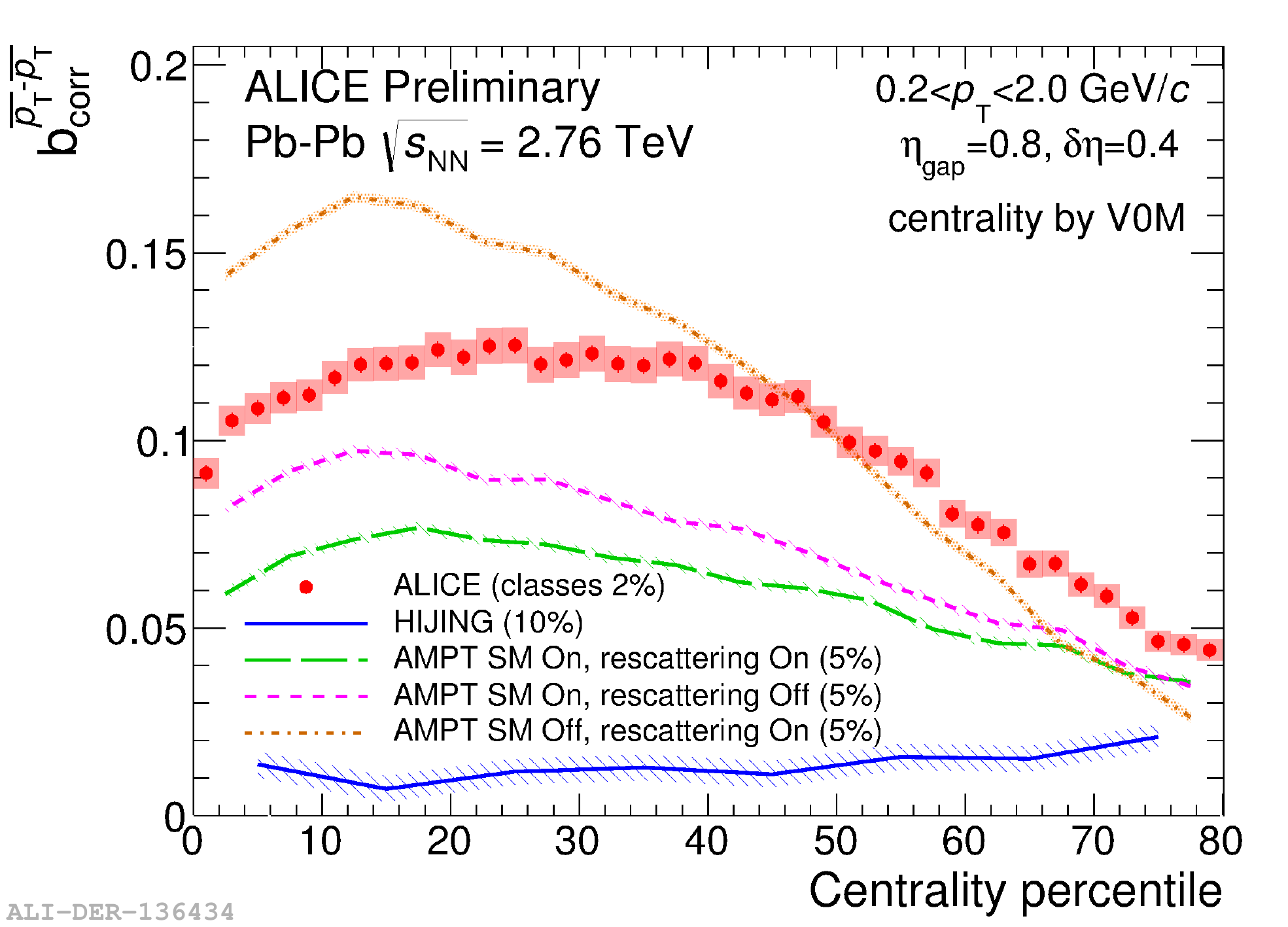}
\end{overpic}
\caption{
Correlation strength $\bcor^{\ptpt}$
between the FB windows  (-0.8, -0.4) and (0.4, 0.8)
in comparison with event generators:
HIJING (blue solid line) and  tunes of the AMPT (dashed lines).
}
\label{fig:ptpt_models_HIJING_AMPT} 
\end{figure}

Figure \ref{fig:ptpt_models_HIJING_AMPT} compares
the FB mean-$\pt$ correlation strength with 
calculations in Monte Carlo generators. 
HIJING demonstrates weak correlations with no dependence on centrality.
Small positive values of $\bcor^{\ptpt}$ in this generator can be attributed to back-to-back jets,
 which hit both $F$ and $B$ windows.
AMPT generally reproduces the shape of the centrality dependence, 
however, 
it does not reproduce the magnitude.
Switching off rescattering or string melting mechanisms 
leads to a rise of   $\bcor^{\ptpt}$, the underlying reasons for this 
need to be investigated further. 
Calculations of the mean-$\pt$ correlations 
in some other event generators are given in
\cite{ptpt_MC_generators_VK}.

\section{Summary}

In summary,
it is shown that the strength of forward-backward correlations 
between multiplicities
heavily depends on the centrality determination procedure
(type of centrality estimator and class width),
therefore, any physics conclusions should be made very carefully.
The FB correlations between mean-$\pt$
have been 
measured for the first time in ALICE in Pb--Pb collisions. 
Correlations of this type 
are robust against volume fluctuations and thus the centrality determination methods,
and, therefore,  provide higher sensitivity to the properties of the initial state and evolution of the medium created in A-A collisions. 
The correlation strength rises from peripheral to mid-central and
drops towards central collisions.
This evolution with centrality is described by some models qualitatively, but not quantitatively. 

\placetextbox{0.88}{0.075}{5}

\section*{Acknowledgements}
This work is supported by the Russian Science Foundation, grant 17-72-20045.

\end{document}